# Fast Ultrahigh-Density Writing of Low Conductivity Patterns on Semiconducting Polymers

Marco Farina[1*], Tengling Ye[2], Guglielmo Lanzani[2], Andrea di Donato[1], Giuseppe Venanzoni[1], Davide Mencarelli[1], Tiziana Pietrangelo[3], Antonio Morini[1], Panagiotis E. Keivanidis[2*]

[1]Dept. of Information Engineering , Università Politecnica delle Marche, Via Brecce Bianche, 60131 Ancona, Italy
[2]Centre for Nano Science and Technology, Fondazione Istituto Italiano di Tecnologia, Via Giovanni Pascoli 70/3,20133 Milano
[3]Dept. of Neuroscience and Imaging, Università "G. d'Annunzio', Chieti, Italy

**The exceptional interest in improving the limitations of data storage, molecular electronics, and optoelectronics has promoted the development of an ever increasing number of techniques used to pattern polymers at micro and nanoscale. Most of them rely on Atomic Force Microscopy to thermally or electrostatically induce mass transport, thereby creating topographic features. Here we show that the mechanical interaction of the tip of the Atomic Force Microscope with the surface of a class of conjugate polymers produces a local increase of molecular disorder, inducing a localized lowering of the semiconductor conductivity, not associated to detectable modifications in the surface topography. This phenomenon allows for the swift production of low conductivity patterns on the polymer surface at an unprecedented speed exceeding 20 $\mu$m s$^{-1}$; paths have a resolution in the order of the tip size (20 nm) and are detected by a Conducting-Atomic Force Microscopy tip in the conductivity maps.**

---

[*] To whom correspondence should be addressed. E-mail: m.farina@univpm.it; pekeivan@iit.it

Multicomponent organic thin films exploiting conjugated polymers are being investigated as candidates for optoelectronic devices[1,2,3,4,5,6]. Among them, poly(3-hexylthiophene (P3HT) / [6,6]-Phenyl-C71-butyric acid methyl ester (PCBM) blends are suitable in photovoltaic applications, with photoconversion efficiencies of up to 5% over small areas[7]. Owing to the interpenetrating microphase-separated internal structures, these devices are usually indicated as bulk heterojunction solar cells. According to the frontier energy levels of these materials (Fig. 1), P3HT is the electron donor and PCBM is the electron acceptor; in solar cells, the P3HT absorbs in the visible part of the spectrum, generating excitons which are dissociated at the donor/acceptor interfaces[5]. In the interpenetrating network, holes move along the P3HT phase and electrons along the PCBM phase, up to collection at the electrodes. Thus, the performance of organic photovoltaic blends is affected by the film morphology, that influences transport properties[8]. Moreover, the intrinsic mobility is heavily related to the molecular order of each component of the blend, and therefore, the nanoscale structure of organic solar cells has been investigated using several different microscopy techniques[9].

Among the techniques used, C-AFM (Conductive- Atomic Force Microscopy) is a scanning probe technique that allows to measure simultaneously the surface topography and the electrical properties of a sample with nanometric resolution. In details, C-AFM provides a way to perform estimations of important parameters, such as local hole and electron mobility[10], and highlighting the relative size of donors and acceptor domains. In C-AFM a sharp metal coated tip is scanned across the sample, and the short range tip-sample interaction is used as feedback control to monitor the tip-sample distance. In its simplest application, the tip is in contact with the sample and the feedback ensures that the tip exerts a constant force. In this mode, the tip also acts as a nanoelectrode, thereby recording the current when the sample is biased with respect to the tip (Fig. 1a). Most importantly, all standard measurements supported by AFM can be carried out simultaneously (friction force –longitudinal and transverse-, force-distance spectroscopy, topography etc).

In view of using polymers in organic electronics, an ever increasing number of techniques used to pattern polymers at micro and nanoscale is being promoted[11], some directly involving AFM devices. While most of these techniques are based on selective removal, oxidation, or deposition of polymers, electrostatic nanolithography based on AFM[12,13] exploits the electrostatically induced mass transport of polymer, thereby creating micrometric topographic features within time scales spanning from milliseconds to seconds.

In this work we show that the mechanical friction of the AFM tip can be used to modify the local molecular order on the surface of several conjugate polymeric semiconductors and blends, and in particular in the prototype photovoltaic organic blend P3HT-[70]PCBM, inducing an immediate reduction of the local carrier mobility. The increase in local disorder is demonstrated by Lateral Force[14,15] images, and modeled in the framework of the Gaussian Disorder Model[16]. The effect results in a non-topographic high resolution patterning of the polymer.

**Results**

**Non-topographic patterning**. It is known that AFM, in contact mode, may easily abrade soft surfaces. However, in several classes of polymers, AFM-induced plastic deformations, which appear as surface nano-ripples, were related to second-order phase transitions (glass-to-rubber)[17].

In the specific case of conjugate polymeric blends P3HT:[70]PCBM, we have found that the phase transition induced by the sliding AFM tip results in locally disordered (amorphous) regions, where the conductivity appears heavily reduced. Interestingly, this happens even when the topographical modification only results in a marginal change in roughness (in the order of 0.1nm rms or less) or there is no detected modification.

In particular, we have spun 100 nm thick P3HT:[70]PCBM films onto ITO, and annealed these samples at different temperatures (ranging from 100 °C to 200 °C), below the melting temperature of pure P3HT, which is ca. 225 °C. Then, a non-topographic "writing" effect was achieved by performing an AFM scan without applying any bias (Figure 2), hence without any electrostatic field or thermal excitation. Indeed, subsequent scans highlighted the previously scanned area as a region with lower conductivity. The size of the patterns (20nm wide) is in the order of the size of the AFM tip, and patterns appear immediately. Owing to this origin, patterns are stable and, in the limit of the polymer deterioration, permanent.

In order to test patterning stability, we have created a feature, which appeared unmodified after five days. We have observed the phenomenon in thin films of several conjugate polymers and blends, such as P3HT, P3HT:[70]PCBM, F8BT:[70]PCBM, blends involving P3HT with different molecular weight, and blends of P3HT:[70]PCBM annealed at different temperatures. Depending on the mobility, on the band structure in our measurement setup and on the thermal history of samples, the detection of the mobility reduction was more or less easy.

The phenomenon was particularly evident and easily detectable for P3HT:[70]PCBM; as discussed in the Supplementary Information (Supplementary Note 2), this is due to a number of different factors; among them, there is likely the low glass transition temperature of P3HT[18] and the relatively high order of P3HT crystallites. The molecular disorder in fact is directly related to the variance of the energy profiles of the Density of States (DOS); in blends there is also an additional effect induced by the amount and the orientation of the dipole moments in the system[19], since interface dipoles are formed at the P3HT/PCBM heterojunctions. A relevant additional factor is that the measured current may heavily depend on the metal used for the tip coating and the conductive substrate. In fact the resulting band structure of the system tip/blend/substrate will occasionally create diode-like junctions. This is for example the case of F8BT:PCBM blend deposited on Indium Tin Oxide (ITO) coated glass and measured by a Platinum coated tip (Fig. 1c), where a potential barrier appears between tip and sample, making difficult to detect the effect of writing (Supplementary Fig. S1).

Figure 3a, reporting a zero bias current map of a region where we had previously driven the C-AFM tip along a vector path to write the word "OK", shows clearly the patterning effect. As anticipated, the topography reported in Figure 3b and in Figure 3d (in 3D), and the error image (Figure 3c) do not show any topographic trace of such a writing, in spite of a relatively regular and flat surface.

The reading process involved in Figures 3a-d is simply a further scan with a lighter set-point, according to what summarized in Figure 2. The AFM tip is driven near the threshold where the tip is no longer in contact with the sample, in order to ensure that no overwriting happens. In annealed P3HT:[70]PCBM some small current was recorded in non-written areas even with nominal zero voltage bias, so that maps of currents showed clearly the writing without the need of bias (Figure 3a is at nominal zero bias). Measurements by an oscilloscope confirm that nominal zero bias corresponds to a couple of mV of offset in the instrument (NT-MDT P47 microscope) with an amount of electrical noise, sufficient to read clearly the stored information.

Owing to the work function of Platinum (Fig. 1b) when the Pt tip serves as an electrode, the J-V curves of the P3HT:[70]PCBM films on ITO are equivalent to those of a hole-only device. In such a configuration low-voltage current maps in P3HT:PCBM are basically hole current images and the best reading is obtained with very low bias; indeed when increasing the bias, the current map shows the sample inhomogeneity, in particular highlighting, as expected[10], either P3HT (bright regions) or [70]PCBM(dark regions) rich domains and partly hiding the written features (Supplementary Figure S2).

By using a soft tip, the CSC17/Pt/AlBS from Mikromasch, having typical force constant 0.15 N/m (range 0.05-0.3 N/m), we were able to estimate from the deflection/distance spectroscopy (Supplementary Figure S3, as example) the minimum interaction allowing a visible writing in the order of 1 nN or less. This value depends also on the annealing temperature of the samples, since the mechanical properties of the surface are also related to the thermal history of sample. Multiple reading can be achieved by using soft tips and very light set-points in the feedback chain. Once writing happens, the change in conductivity is slightly modulated within 2-3 nN of force (i.e. a higher contrast in the current maps may be obtained by slightly increasing the writing set-point, hence the load). A further increasing of the tip pressure does not improve the writing contrast, since the low-bias conductivity is dramatically reduced, but can only lead to damages on the sample surface and to lithographic changes (Supplementary Figure S4).

A moderate bias does not affect the writing effect. This is shown in Figure 4a, reporting the current map in a region where three squares were written at different biases (-0.5, 0, or +0.5

V). On the other hand, a large bias, in the order of several volts, may induce a sharp increase of current up to several nA, triggering the melting and (relatively slow) shaping effects. We believe that this is the effect described by Zaniewski[13], where the biasing electric field induces a polymer ordering and a corresponding *increase* of conductivity. In our case, the writing corresponds to a lowering of the polymer conductivity, as shown in Figure 4b, where average I-V curves (10 measurements) -along with standard deviation- are evaluated inside and outside a written area in argon saturated atmosphere. As matter of fact, scans were performed either in air or in argon saturated atmosphere. While quantitative measurements of current-voltage characteristics are more reliable in controlled atmosphere, the writing process was unaffected, ruling out a possible impact of the room moisture -inducing a water meniscus between tip and sample- or tip-induced oxidation.

For the sake of completeness, samples at different annealing temperatures were characterized in terms of IV curves; the behavior of the curves confirms the expected beneficial effect of the annealing with respect to carrier mobility (Supplementary Figure S5, measurement in air), since annealing induces phase separation between P3HT and [70]PCBM and crystallization of the ordered P3HT domains.

**Modeling.** The writing phenomenon is well accounted for in the framework of the Extended Gaussian Disorder Model (EGDM) proposed in ref.[16]. Such a model is based on the numerical solution of the master equation for hopping transport, assuming a disordered energy landscape with a Gaussian Density Of States, and provides estimation of the mobility as function of the electric field, of the carrier density and of the temperature (see Methods for details; Supplementary Figure S6 for temperature behavior).

We have also obtained average –field independent- mobility values[20] ranging from 0.0001 (as spun sample) to 0.012 (annealed) cm$^2$ V$^{-1}$ s$^{-1}$ (Supplementary Figure S7, Supplementary Note 1). In such calculations a slightly increased effective contact area, accounting for the surface spreading of the current, was used for the AFM tip, overcoming a potential over-estimation of the actual value[21] of mobility. Consequently, the obtained values agree quite well to the values reported in the literature[10] (Supplementary figure S7). Most importantly, we always observe a significant reduction in mobility in written regions (for example from the original value of 0.012 –unwritten- to 0.005 cm$^2$ V$^{-1}$ s$^{-1}$) which is well described by a slight broadening of the width of the Gaussian density of states, ranging in a fraction of eV, consequent to the induced molecular disorder.

In order to experimentally ascertain that the reduction in conductivity can be related to the increase in the molecular disorder of the polymeric surface, we have used the Lateral Force (LF) (or Friction) imaging. Basically, during the scan, the twisting of the AFM cantilever is simultaneously mapped. Lateral Force and its variant, the Transverse Shear Microscopy, have been demonstrated to be powerful tools to investigate crystallographic features of surfaces[14,15]. In particular, the amorphous versions of any polymer feature a larger number of molecular-scale degrees of freedom than their crystalline counterparts. Thus, the translational kinetic energy of a sliding AFM tip is transferred to a large number of molecular motions, and modes for this to occur are larger in amorphous domains because of the greater molecular freedom. The consequence is that a region with lower molecular order will appear as a brighter region in LF scans.

This is actually what happens in our case; Figure 5 shows a square that was written in P3HT:[70]PCBM, appearing not only in current images (5a), but also as a bright region in LF image (5b) and being invisible in topography (5c). The effect is also represented in 3D mode by writing the acronym "UNIVPM" in current (5d), LF (5e) and topography (5f). We were not able to see the same effect in some other materials, such as PDBTTT and the system PDBTTT:PDI, an alternative in current research for organic photovoltaic cells, as demonstrated in Supplementary Figures S8 and S9 and discussed in the Supplementary Note 2.

Images were also obtained in semicontact mode (See Supplementary Figures S10, S11, Supplementary Note 2) to further investigate surface modifications, and highlighting the fibril-like P3HT crystals in the surface, agreeing with previously reported in TEM microscopy study[22,23]. Note that in this scanning mode the AFM tip does not "write", demonstrating that friction triggers the creation of the low molecular order layer. In this mode, also phase images are available (Supplementary Figure S12); even though phase images are always difficult to interpret, being critically dependent by several concurrent parameters, they also show to some extent the existence of the amorphous layer.

## Discussion

In this work we demonstrate a novel "nano-writing" effect for several semiconductor polymers and blends, particularly efficient in P3HT:[70]PCBM films, where a sliding AFM conductive tip induces local changes in the polymer conductivity, not associated to visible topographic changes. The width of the written paths is in the order of the curvature radius of the AFM tip; patterns appear almost immediately and their creation does not need either biasing or heating. By using LF imaging, we demonstrate that this is due to a reduction of the molecular order, whose impact is enhanced by the interpenetrating microphase-separated structures of the polymeric heterojunction, resulting in an increase of the electrical resistivity. The effect is well described as an increase in the Gaussian variance of the Density Of States in the framework of a Gaussian Disorder Model.

High resolution, high write speed, low cost and low power consumption suggest this as a viable technique to be explored for several applications in polymer electronics. In particular, the technique could be optimized for the ultra-high density non-volatile data storage; in fact by writing a feature of 20nm as a basic bit, one would obtain a theoretical maximum data density of 1.6Tb in$^{-2}$, exceeding the current technological limit, estimated in the range of 1 Tb in$^{-2}$.

Of course, this practical application would need an optimization of the polymeric blend in several directions, trying to reduce all sources of noise when reading bits (hence finding the best bias for reading, reducing the size of local conductivity inhomogeneity etc) and looking for the blend preparation providing the optimal "writeability", namely best contrast between written and unwritten areas. Significant work should be done to assess long term stability (months or more) and to develop a dedicated system exploiting our concept to store a significant amount of data.

Nevertheless, the writing speed and the data density may represent a technological advantage with respect to other techniques being investigated, such as the electrostatic nanopatterning. Moreover very modest forces are needed and no thermal excitation, making the principle viable for low power consumption systems. Importantly, the organic composite layer is inexpensive, easy to prepare, and to scale for very large areas; these features have to be kept in mind when considering challenges in terms of cost of recent promising techniques, as e.g. in the work by Cen et al[24], where the Authors were able to exploit metal-insulator quantum phase transition at the interface between two insulating oxides to write conducting paths 3nm wide.

From a general point of view, our writing procedure may resemble to a standard mechanical indentation process, hence similar to the one proposed by Kim and collaborators[25], who reported that the baroplastic properties of block copolymers may be utilized to write bits (up to 1 Tb in$^{-2}$) as depressed topographic features by an AFM tip. Incidentally, block copolymers are being actively investigated for mass memory device[26], mostly owing to their self-assembly capabilities[27]. However, in our case the physical process is completely different, being induced by the sliding tip, and reading is not related to the detection of topographic depressions, but only to changes in the conductivity. This observation also suggests further possible advantages, since reading of patterns is almost unrelated to topographic imperfections.

As matter of fact, an efficient AFM feedback is able to drive the tip on grains and surface imperfections and, as shown for example in Figure 2, the topography is generally not correlated with the conductivity map. In this way, requirements for surface flatness are mitigated. Moreover, the maximum achievable density of stored data would only be related to the size of the C-AFM tip, which in principle can be reduced to a few nanometers.

**Methods**

**Preparation of samples.** Two sets of P3HT alone and P3HT:[70]PCBM were prepared, namely a first set (Regioregularity: 95.7%; molecular weight (Mw): 65200 gr/mole; polydispersity index: 2.2; Mg content: 6 ppm; Ni content: <1 ppm) and a second set (Regioregularity: 95.2%; molecular weight (Mw): 36600 gr/mole; polydispersity index: 2.0; Mg content: 24 ppm; Ni content: 10 ppm). Annealing was performed on a high-accuracy hotplate at 100°-140°-160°- 200°C for 15' in ambient and rapidly quenched back to room temperature. ITO had a nominal surface resistivity of 15 Ohm/Sqr. For both set of samples, a corresponding sample with an additional layer of PEDOT:PSS (47nm) interposed between ITO and P3HT:PCBM was prepared. No significant difference in the writing behavior was observed, in spite of a significant reduction in the surface roughness.

Additional polymers were also prepared: F8BT:PCBM70 (1:1) as spun, on ITO, PDBTTT:PDI blend and PDBTTT alone.

**C-AFM measurements.** Measurements were performed either on air or saturated Argon atmosphere, by an NT-MDT Solver Pro P47 AFM, equipped with head for conductive, contact, semicontact and tapping measurements. In particular the C-AFM measurements were realized by means of measuring head AU006. This is an adjustment unit for contact and resonant AFM, including the capability to measure current through conductive cantilever; the tip is kept at ground potential while the sample is biased by a controlled voltage generator (Figure 1). Semicontact images were obtained by the same microscope.

Several kinds of tips have been used: images shown in Figure 1 have been obtained by the NSC19/Pt/no Al from Mikromasch. This is a silicon tip coated by 14 nm of Pt, featuring a curvature radius less than 24 nm and a typical force constant of 0.63 N/m (ranging between 0.17 and 1.7 N/m). In order to evaluate the minimum interaction force, as detailed in the Supplementary Information, we have used a softer tip, the CSC17/Pt/AlBS from Mikromasch, having a typical force constant 0.15 N/m (range 0.05-0.3 N/m). Images in Figure 4 have been obtained by a tip NSC18/Pt, force constant 2.8 N/m (range 1.2-5.5 N/m), radius <30nm. Remaining scans were performed by NSC19-Ti/Pt, having the same features of the first set of tips, but exploiting an alloy of Titanium/Platinum as coating. Some additional tests (results not included) were performed by NSG03/Co/5 from NT-MDT, a tip featuring cobalt coating (typical k=1.1N/m, range 0.5-2.2N/m), used to investigate and exclude the possible impact of the kind of conducting coating in the writing process.

**Extended Gaussian Disorder Model [16].** By numerically solving the master equation for hopping transport, assuming a disordered energy landscape with a Gaussian density of states (DOS), Pasveer and coauthors[16] have successfully determined the dependence of the charge-carrier mobility for semiconductor polymers on temperature, carrier density, and electric field. The results of the numerical solution of the master equation have been parameterized[16], providing an approximation for the mobility:

$$\mu(T, p, E) \approx \mu(T, p) f(T, E)$$

Where

$$\mu(T, p) = \mu_0(T) \exp\left[\frac{1}{2}(\hat{\sigma}^2 - \hat{\sigma})(2pa^3)^\delta\right]$$

$$\mu_0(T) = \mu_0 c_1 \exp[-c_2 \hat{\sigma}^2]$$

$$\delta \equiv 2 \frac{\ln(\hat{\sigma}^2 - \hat{\sigma}) - \ln(\ln(4))}{\hat{\sigma}^2}$$

$$\mu_0 = \frac{a^2 \nu_0 e}{\sigma}$$

$$f(T, E) = \exp\left\{0.44(\hat{\sigma}^{3/2} - 2.2)\left[\sqrt{1 + 0.8\left(\frac{Eea}{\sigma}\right)^2} - 1\right]\right\}$$

$$\hat{\sigma} = \frac{\sigma}{k_B T}$$

In the above set of equations, $c_1 = 1.8 \times 10^{-9}$ and $c_2 = 0.42$. Moreover $\sigma$ is the width of the Gaussian density of states (in eV), $\nu_0$ is the intrinsic hopping rate, $a$ is an effective lattice constant, $p$ the carrier density, $e$ the electron charge, $k_B$ the Boltzmann' constant and $E$ the electric field. In this model, $a$ and $\nu_0$ are fitting parameters, along with the width of the Gaussian DOS. Generally the carried density $p$ is function of the position $x$. In order to obtain the low-voltage IV curves the above results are inserted in the expression

$$J = e\, p(V) \mu(T, E, p) E(L), \text{ with } J = I/(\pi r^2);$$

here $J$ is the current density, $r$ is an estimated radius of the current-density spot, $d$ is the thickness of the sample, and $\varepsilon$ is the dielectric constant. We still need the density of carriers $p$, which in turn is related to the electric field by the Gauss equation

$$\frac{dE}{dx} = \frac{e\,p}{\varepsilon}$$

In ref.[16] the above equations are solved numerically. In our case, we have a thin film (L=100-200nm); in order to have a simple, closed form, approximation, we can assume the charge density to be spatially uniform, so that the electric field is a linear function of space. In this way, the obtained value for the charge density can be seen as a space averaged effective value. In the limit of such an assumption, by integrating the electric field across the film to obtain the applied voltage, we can write $p$ as function of the applied voltage

$$p(V) = \frac{2V\varepsilon}{eL^2}$$

In order to compare the experimental data with the theory, measurements for P3HT:PCBM were performed at relatively low voltages, to avoid injection of electrons and measuring hole-only mobility. At low voltage, only positive carriers contribute to the current: in fact, basing on the band structure shown in Figure 1b, the ITO Fermi-level lies just above the HOMO energy level of P3HT and, hence, holes see a very low energy barrier, easily overcome even with small biases. In particular, for positive (negative) voltages, holes are injected from the ITO-electrode (Pt-tip) to the Pt-tip (ITO-electrode), through the P3HT valence energy-levels. At higher voltages, both positive and negative, also electrons can partially contribute to the current, as they gain energy and can be injected over the forbidden gaps (of about 2 eV) of P3HT and PCBM. Figure 6 shows the comparison between the EGDM results (from the above equations) and data recorded at room temperature for the blend involving the higher molecular weight P3HT. EGDM fitting was obtained by selecting σ=0.11eV for the clean area and σ=0.116eV for the written area.

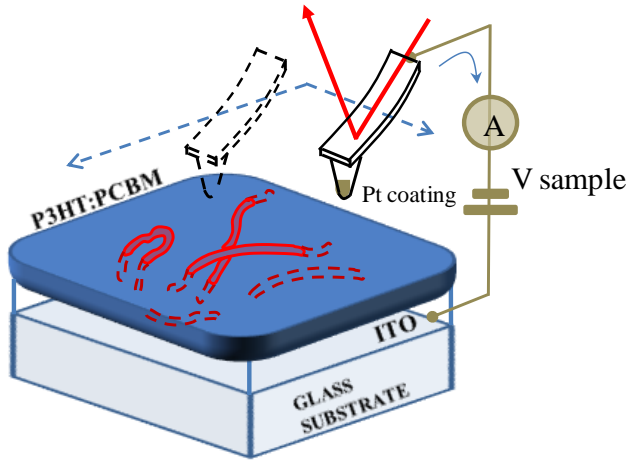

a

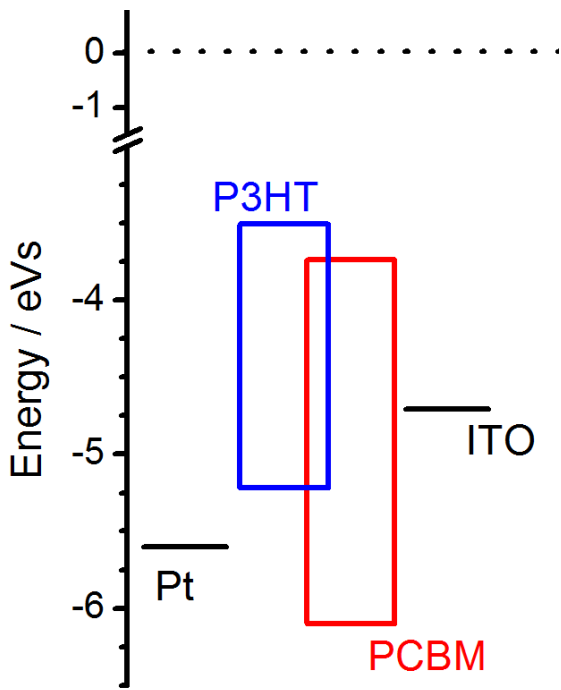

b

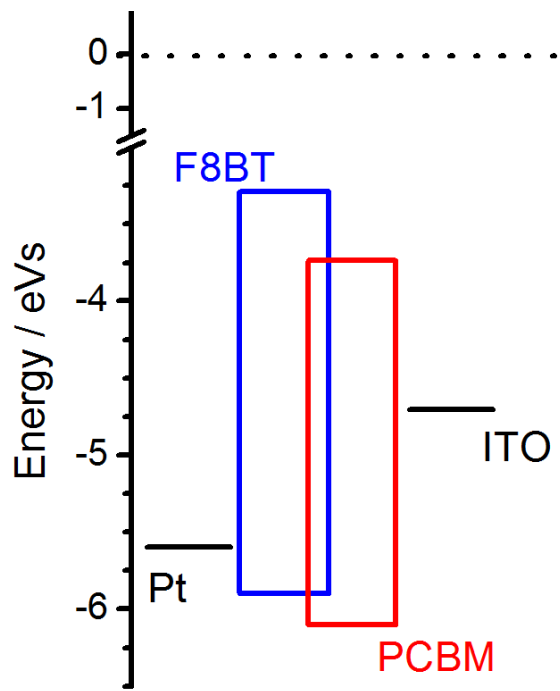

c

**Fig. 1: Conductive Atomic Force measurement.** Panel **a** shows our measurement setup; panel **b** highlights HOMO and LUMO energy levels of P3HT and PCBM compared to the work functions of Pt and ITO (referenced with respect to a vacuum)[10] while panel **c** HOMO and LUMO levels for F8BT:PCBM. Note that in our set up the AFM tip was grounded and the ITO was negatively biased in reverse bias mode or positively bias in the forward bias mode. Accordingly, the measurement set up allowed to record basically hole-based currents in P3HT:PCBM. In fact, in the low reverse bias regime electron injection from ITO to the LUMO levels of P3HT and PCBM is negligible due to the very high energy barrier, unless electron tunneling occurs. Moreover, hole injection from the Pt to the HOMO of P3HT is favored, resulting in the hole-current of the device.

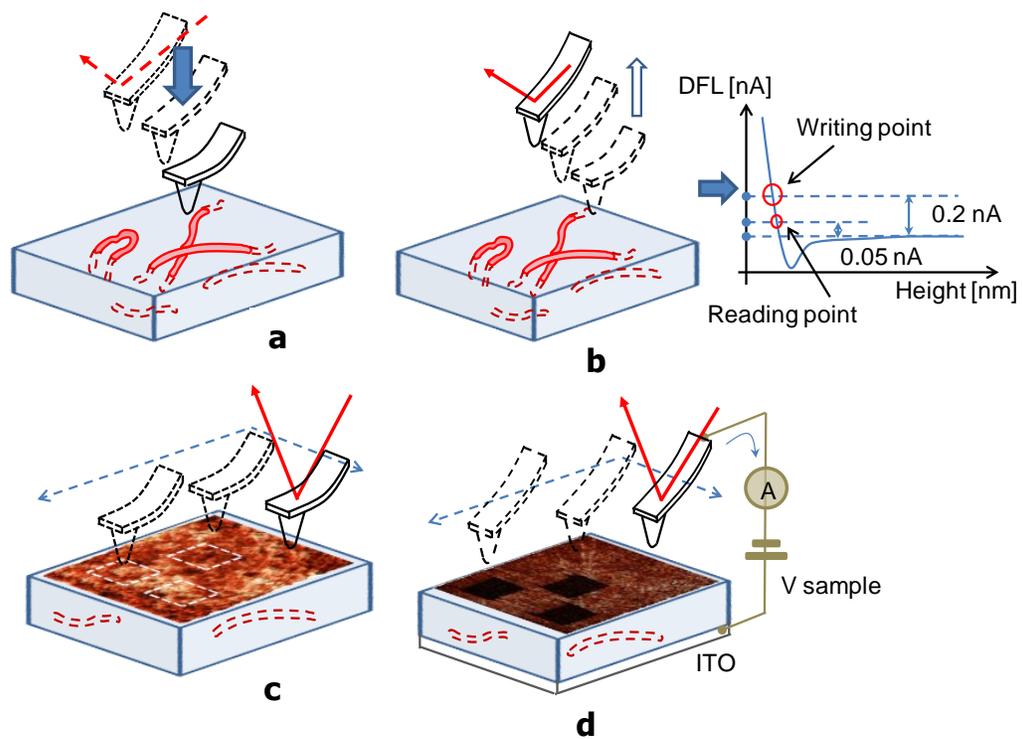

**Fig. 2: Description of the sequences followed in the writing/reading process.** (Panel **a**) the tip is approached to the surface in contact mode; (Panel **b**) definition of the writing and reading points through the spectroscopic curve DFL vs HEIGHT: the reading point is as close as possible to the limit where the tip stops interacting with the surface; (Panel **c**) scanning the surface in contact mode (writing), (Panel **d**) scanning the surface in spreading resistance mode (reading): reduction of the conductivity of written patterns is clearly visible.

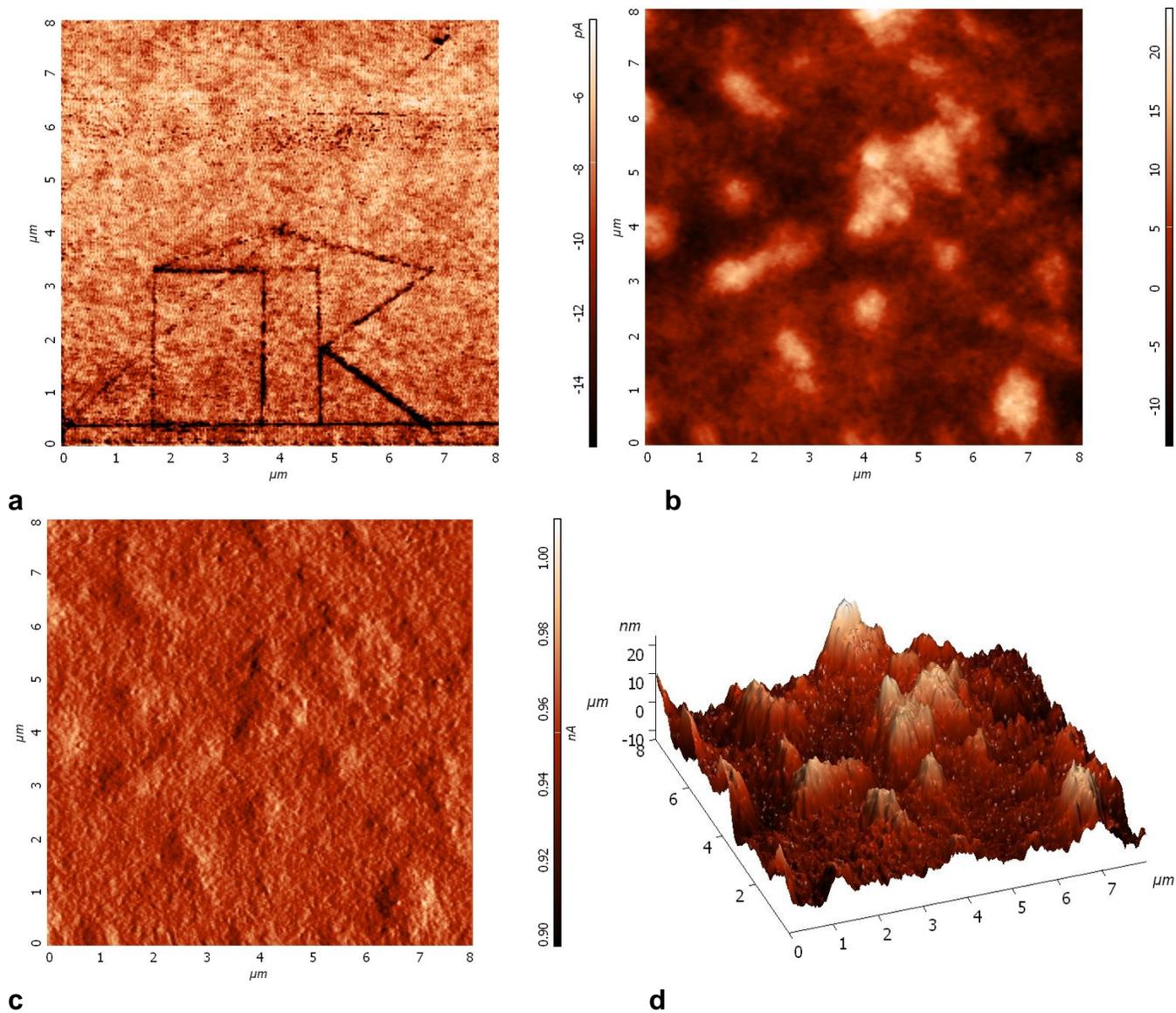

**Fig. 3: Non topographic writing.** Panel **a**: Current map showing the pattern "ok" that we have created by simply moving the AFM tip in contact with P3HT:[70]PCBM annealed at 200°C (note: the current map has a negative offset) ; Panel **b** shows simultaneous topography (corrected for flattening) where there is no trace of the writing. Panel **c** shows deflection (error) image, where possible scratches would be easier to detect. Panel **d** shows topography in 3D images displaying no correspondent topographic effect.

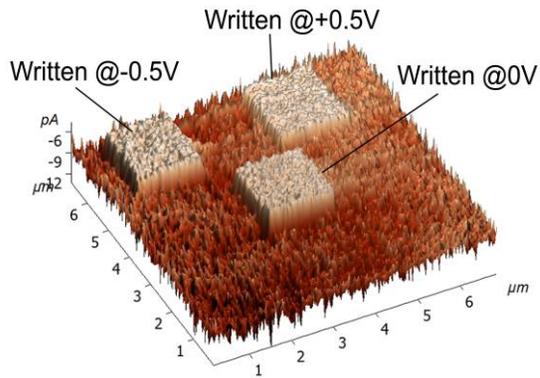 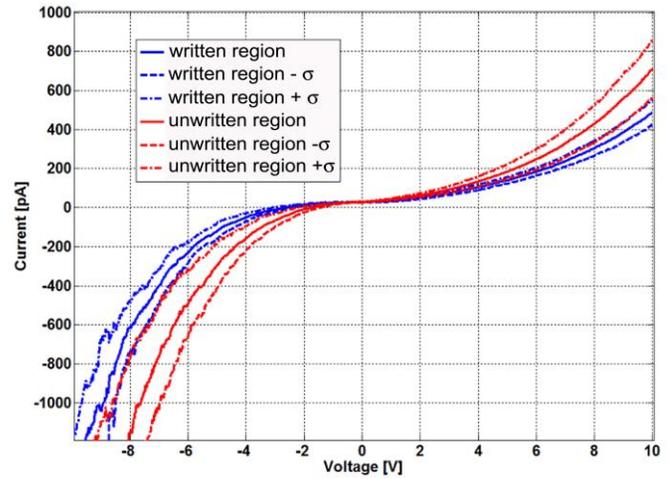

**a**  **b**

**Fig. 4: Effect of biasing of the AFM tip**. Panel **a** displays a current map (image reverted for clarity) showing three lower conductivity squares, created by performing three writing with different biases (P3HT:[70]PCBM annealed at 100°C); apparently, different biases did not affect the writing process. Panel (**b**): average (and standard deviation) I-V curves for written and unwritten regions (Sample P3HT:[70]PCBM annealed at 100°C). The written region results in reduction of conductivity.

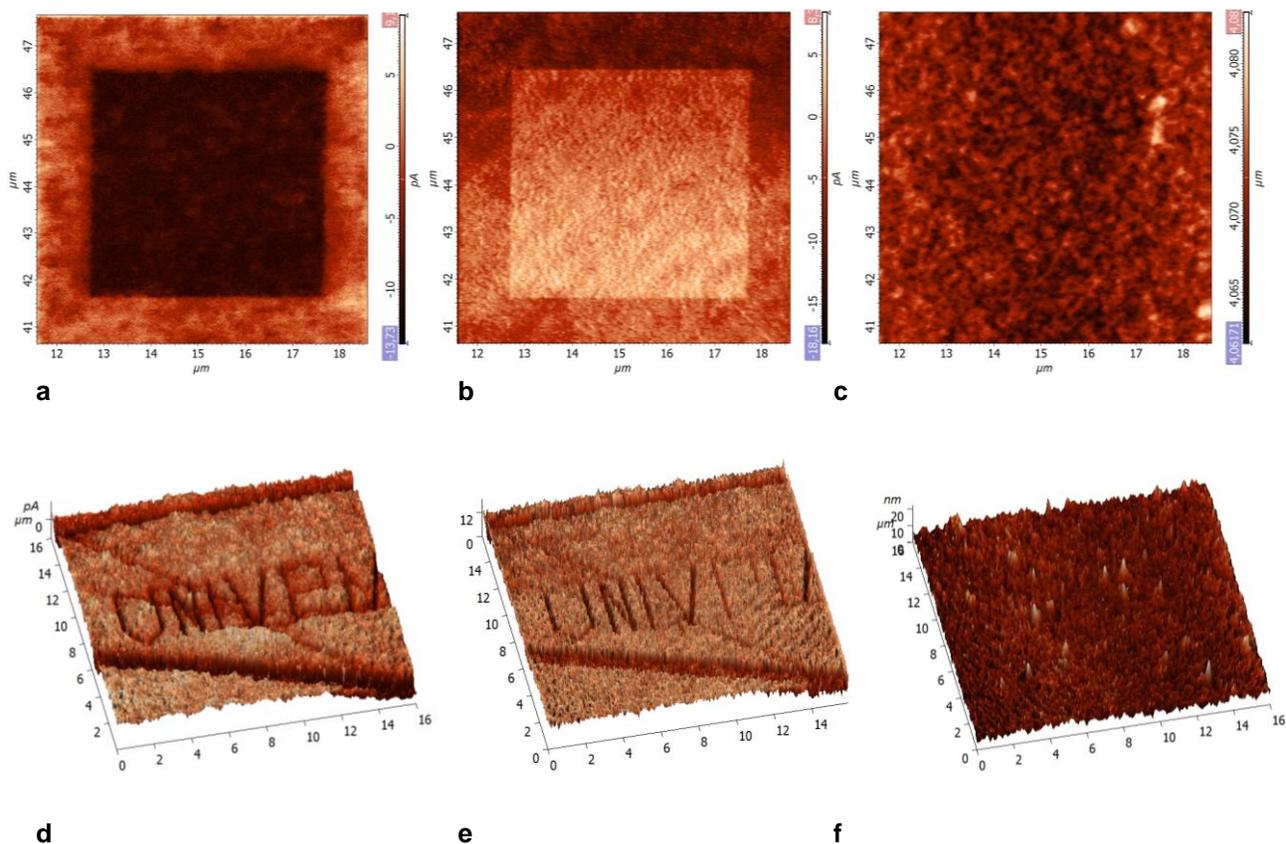

**Fig. 5**: **Comparison between current maps, lateral force and topographic images.** A square, 5μm of side, was written; reading is performed in an area 7μmx7μm in P3HT:[70]PCBM (1:1) annealed at 100°C for 15mins. (**a**) Current map (0.5V) (**b**) Lateral Force image: the written square appears as a region with increased molecular disorder (**c**) topography: no topographic change is visible. The above images were obtained simultaneously. Also, the acronym "UNIVPM" is written and represented in 3D: (**d**) current map, (**e**) Lateral force (inverted for clarity) and (**f**) topography.

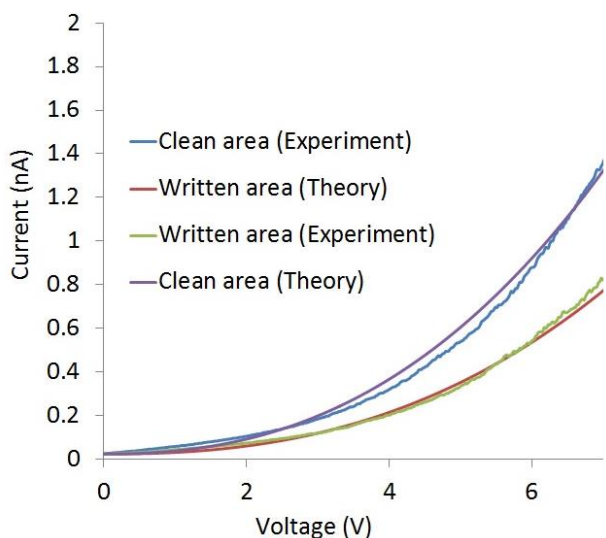

**Fig. 6: EGDM model.** Current-Voltage characteristic as for P3HT:[70]PCBM annealed at 100 °C (blend involving higher molecular weight P3HT) for unwritten region and written region in same experiment, recorded at room temperature (T=300K) and comparison with EGDM model. In the clean area the width of the Gaussian DOS is σ=0.11eV, while σ=0.116eV in the written region. Experimental curves are average over 5 measurements.

# References and Notes

1. Yu, G., Gao, J., Hummelen, J. C. & Heeger, A. J. Polymer Photovoltaic Cells: Enhanced Efficiencies via a Network of Internal Donor-Acceptor Heterojunctions. Science 270, 1789–1791 (1995).

2. Park, S. H. et al. Bulk heterojunction solar cells with internal quantum efficiency approaching 100%. Nature Photonics 3, 297–302 (2009).

3. Gong, X, Soci, C. , Yang, C., Heeger, A.J. & Xiao S. Enhanced electron injection in polymer light-emitting diodes: polyhedral oligomeric silsesquioxanes as dilute additives. J. Phys..D: Appl. Phys. 39, 2048-2052 (2006).

4. Chiesa, M. et al. Correlation between surface photovoltage and blend morphology in polyfluorene-based photodiodes. Nano letters 5, 559–563 (2005).

5. Halls, J. J. M., et al. Efficient photodiodes from interpenetrating polymer networks Nature 376, 498-500 (1995)

6. Goffri, S. et al. Multicomponent semiconducting polymer systems with low crystallization-induced percolation threshold. Nature Materials 5, 950 - 956 (2006)

7. Kim, J. Y. et al. New Architecture for High-Efficiency Polymer Photovoltaic Cells Using Solution-Based Titanium Oxide as an Optical Spacer. Advanced materials 18, 572–576 (2006).

8. Veldman, D. et al. Compositional and electric field dependence of the dissociation of charge transfer excitons in alternating polyfluorene copolymer/fullerene blends. J Am Chem Soc.130, (24), 7721-35, (2008)

9. Chen, W., Nikiforov, M. P. & Darling, S. B. Morphology characterization in organic and hybrid solar cells. Energy & Environmental Science 5, 8045 (2012).

10. Dante, M., Peet, J. & Nguyen, T.-Q. Nanoscale Charge Transport and Internal Structure of Bulk Heterojunction Conjugated Polymer/Fullerene Solar Cells by Scanning Probe Microscopy. Journal of Physical Chemistry C 112, 7241–7249 (2008).

11. Nie, Z. & Kumacheva, E. Patterning surfaces with functional polymers. Nature Materials 7, 277–290 (2008).

12. Lyuksyutov, S. F. et al. Electrostatic nanolithography in polymers using atomic force microscopy. Nature Materials 2, 468–472 (2003).

13. Zaniewski, A. M., Loster, M. & Zettl, A. A one-step process for localized surface texturing and conductivity enhancement in organic solar cells. Applied Physics Letters 95, 103308–103308 (2009).

14. Kalihari, V., Haugstad, G. & Frisbie, C. D. Distinguishing Elastic Shear Deformation from Friction on the Surfaces of Molecular Crystals. Physical Review Letters 104, (2010).

15. Kalihari, V., Tadmor, E. B., Haugstad, G. & Frisbie, C. D. Grain Orientation Mapping of Polycrystalline Organic Semiconductor Films by Transverse Shear Microscopy. Advanced Materials 20, 4033–4039 (2008).

16. Pasveer, W. et al. Unified Description of Charge-Carrier Mobilities in Disordered Semiconducting Polymers. Physical Review Letters 94, (2005).

17. Schmidt, R. H., Haugstad, G. & Gladfelter, W. L. Scan-Induced Patterning in Glassy Polymer Films: Using Scanning Force Microscopy To Study Plastic Deformation at the Nanometer Length Scale. Langmuir 19, 898–909 (2003).

18. Zhao, J. et al. Phase diagram of P3HT/PCBM blends and its implication for the stability of morphology. The Journal of Physical Chemistry B 113, 1587–1591 (2009).

19. Young, R. Dipolar lattice model of disorder in random media analytical evaluation of the gaussian disorder model. Philosophical Magazine Part B 72, 435–457 (1995).

20. Mott, N. F. , Gurney, R.W.  *Electronic Processes in Ionic Crystals*, Oxford University Press, London (1948)

21. Reid, O. G., Munechika, K. & Ginger, D. S. Space Charge Limited Current Measurements on Conjugated Polymer Films using Conductive Atomic Force Microscopy. Nano Letters 8, 1602–1609 (2008).

22. Yang, X. et al. Nanoscale Morphology of High-Performance Polymer Solar Cells. Nano Letters 5, 579–583 (2005).

23. Zen, A. et al. Effect of Molecular Weight on the Structure and Crystallinity of Poly(3-hexylthiophene). Macromolecules 39, 2162–2171 (2006).

24. Cen, C. et al. Nanoscale control of an interfacial metal–insulator transition at room temperature. Nature Materials 7, 298–302 (2008).

25. Jo, A., Joo, W., Jin, W.-H., Nam, H. & Kim, J. K. Ultrahigh-density phase-change data storage without the use of heating. Nature Nanotechnology 4, 727–731 (2009).

26. Kim, J. K., Yang, S. Y., Lee, Y. & Kim, Y. Functional nanomaterials based on block copolymer self-assembly. Progress in Polymer Science 35, 1325–1349 (2010).

27. Bates, C. M. et al. Polarity-Switching Top Coats Enable Orientation of Sub-10-nm Block Copolymer Domains. Science 338, 775–779 (2012).

## Acknowledgments

We wish to thank Dr. Alexander Tselev, *Center for Nanophase Materials Sciences*, Oak Ridge National Laboratory, Dr. Tatiana Da Ros, *Dept. of Chemical and Pharmaceutical Sciences*, University of Trieste, Dr. Francesca Maria Toma, Department of Chemistry, University of California Berkeley and Prof. Oriano Francescangeli, *Dipartimento di Scienze e Ingegneria della Materia, dell'Ambiente ed Urbanistica*, Università Politecnica delle Marche, for reviewing the manuscript. We also thank Mr. Stefano Pace for running some of the scans. P.E.K. acknowledges the financial support of an Intra European Marie Curie Fellowship (FP7-PEOPLE 2011-IEF, project DELUMOPV).

## Author Contributions

M.F. and A.D.D. performed the measurements, discovered the effect and processed the data; P.E.K. studied, developed and optimized the samples and conceived the experiments; T.Y. has prepared the samples; P.E.K., G.L., A.M., D.M. and G. V. helped in interpretation of data; M.F. proposed an explanation of the phenomenon; M.F., A.D.D. and T.P. prepared the manuscript; all authors reviewed the manuscript.

## Competing financial interests

The authors declare that they have no competing financial interests.